\documentclass[aps,superscriptaddress,showpacs,nofootinbib,eqsecnum]{revtex4} 
                   
\usepackage{amssymb}  \usepackage{epsf} \usepackage{eucal} \usepackage{epsfig}
\usepackage{graphicx}

\usepackage{graphicx}

\usepackage{amsmath}

\newcommand{\be}{\begin{equation}}
\newcommand{\ee}{\end{equation}}
\newcommand{\bea}{\begin{align}}
\newcommand{\eea}{\end{align}}

\begin{document}

\title{Shear viscous effects on the primordial power spectrum from
  warm inflation} 

\author{Mar Bastero-Gil} \email{mbg@ugr.es} \affiliation{Departamento de
  F\'{\i}sica Te\'orica y del Cosmos, Universidad de Granada, Granada-18071,
  Spain}

\author{Arjun Berera} \email{ab@ph.ed.ac.uk} \affiliation{SUPA, 
School of Physics
and Astronomy, University of Edinburgh, Edinburgh, EH9 3JZ, United Kingdom}

\author{Rudnei O. Ramos} \email{rudnei@uerj.br} \affiliation{Departamento de
  F\'{\i}sica Te\'orica, Universidade do Estado do Rio de Janeiro, 20550-013
  Rio de Janeiro, RJ, Brazil}

\begin{abstract}
 
We compute the primordial curvature spectrum generated during warm
inflation, including shear viscous effects. The primordial spectrum is
dominated by the thermal fluctuations of the radiation bath, sourced
by the dissipative term of the inflaton field. The dissipative
coefficient $\Upsilon$, computed from first principles in the
close-to-equilibrium approximation, depends in general on the
temperature $T$, and this dependence renders the system of the linear
fluctuations coupled.  Whenever the dissipative coefficient is larger than the
Hubble expansion rate $H$, there is a growing mode in the fluctuations
before horizon crossing. However, dissipation intrinsically means
departures from equilibrium, and therefore the presence of a shear
viscous pressure in the radiation fluid. This in turn acts as an extra
friction term for the radiation fluctuations that tends to damp the  
growth of the perturbations. Independently of the $T$ functional
dependence of the dissipation and the shear viscosity, we find that
when the shear viscous coefficient $\zeta_s$ is larger than $3
\rho_r/H$ at horizon crossing, $\rho_r$ being the radiation energy
density, the shear damping effect wins and there is no growing mode in
the spectrum.    

\end{abstract}

\pacs{98.80.Cq, 98.80.Es, 05.70.???a}

\maketitle

\section{Introduction}

Cosmological observations, and in particular Cosmic Microwave
Background (CMB) measurements \cite{COBE,CMB,WMAP}, are
consistent with a nearly gaussian and practically scale invariant
spectrum of primordial perturbations, as predicted by the inflationary
models \cite{inflation, Linde:1983gd}. An early period of inflation also
accounts for the inferred flatness of the Universe, and provides a
solution to the horizon problem. This makes inflation a robust
candidate to account for the early evolution of our
universe. Inflationary predictions are characterized by the spectral
index of the primordial spectrum, its tensor contribution, and the
level of non-gaussianity. Present CMB data however sets at
most an upper limit on the level of the tensor contribution and
non-gaussianity \cite{WMAP}, and it is not yet able
to discriminate among the different implementations and models of 
inflation. This situation is expected to change with the next
generation of CMB experiments, like ESA's Planck surveyor satellite
\cite{planck}, which will further improve our knowledge of the cosmological
parameters.      

Inflation in brief is no more than an early period of accelerated
expansion. In the standard picture of inflation, denoted as cold
inflation, the universe rapidly 
supercools, and inflation should be followed by a  
reheating period, during which the inflationary vacuum energy is
converted into radiation. For reheating to take place, the inflaton field has to
couple to other degree of freedom, such that it can decay into light,
relativistic degrees of freedom that thermalize and provide the
radiation bath \cite{reheating}. During cold inflation one assumes that those couplings
play no role during the accelerated expansion. The alternative, called
warm inflation \cite{bf1,wi}
(for earlier related work see \cite{prewarm}), assumes on the contrary that
those coupling can lead to non-negligible dissipative effects, and radiation
production can occur concurrently with the inflationary expansion. 
Both background evolution and inflaton fluctuations are modified with
the inclusion of an extra friction term $\Upsilon \dot \phi$
accounting for the transfer of energy between the inflation and the radiation. 
The dynamics of the fluctuations are now governed by a
Langevin equation including  a noise force term from the influence
of the radiation fluctuations into the inflaton field 
\cite{bf1,langevin,warmdeltap,BR1,langevin2}. Thermal fluctuations in
the radiation are transfered to the inflaton and become the main
source of primordial fluctuations \cite{bf1,warmdeltap,warmpert}.  

The dissipative coefficient can be computed from first principles in
quantum field theory, within an adiabatic approximation. The two-stage
interaction configuration proposed in \cite{BR1} has been
shown to lead to a large enough dissipative coefficient while keeping
the corrections to the inflationary potential under control and
allowing a period of slow-roll inflation
\cite{br05,Hall:2004zr,Moss:2008yb}. 
For the two-stage mechanism, the inflaton field 
couples to a heavy catalyst field, and the latter in
turn couples to light degrees of freedom. During the motion of the
background inflaton, it excites the catalyst fields which then decay
into light fields \cite{Berera:2008ar}. 
Using this mechanism, the first calculation of the
dissipative coefficient within the close-to-equilibrium approximation
was done in \cite{mossxiong}, leading to a temperature dependent
dissipative coefficient. In the low-temperature regime, when the mass
of the heavy catalyst field is much larger than $T$, one has $\Upsilon \propto
T^3$, while in the high-temperature regime,
when $T$ is above the heavy catalyst field
mass, depending on type of interaction the  dissipative coefficient
becomes linear with $T$ \cite{BasteroGil:2010pb}
or goes goes like the inverse of $T$ \cite{BGR}. All high-$T$ models
suffer in general from very large thermal corrections which spoil the
flatness of the potential \cite{BGR,Yokoyama:1998ju},
with only a few exceptions \cite{warmdeltap,Berera:1998px}.
However, viable
models of warm inflation  have been studied in the low-$T$ regime 
\cite{BasteroGil:2009ec,Zhang:2009ge,joao}.  

The temperature dependence of the dissipative coefficient induces the
coupling of the field and radiation fluctuations as shown in
\cite{mossgraham}. Previous studies of the primordial spectrum of
perturbations in warm inflation \cite{warmpert} did take into account
the influence of the thermal fluctuations on the field through the
noise term, but  not the coupling through the dissipative term
itself. In \cite{mossgraham} it was shown that for positive power of $T$
in $\Upsilon$, and when $\Upsilon$ dominates over the Hubble expansion
rate, this coupling induces a growing mode in the fluctuations before
horizon crossing, enhancing by several order of magnitude the
amplitude of the primordial perturbations with respect to previous
calculations. In the calculations of the primordial spectrum
typically the radiation bath is treated as a perfect fluid, with an
equilibrium pressure $p_r \simeq \rho_r/3$, where $\rho_r$ is the radiation
energy density, an approximation valid in the close-to-equilibrium
regime required for the calculation of the dissipative coefficient to
hold. However, even in that regime,  the radiation bath is expected to
depart from  an ideal fluid as a consequence of the constant
production of radiation particles from  the background field
dissipation.  Imperfect fluids have dissipative
effects that can be parameterized in terms of shear and bulk viscosity
coefficients, and a heat flow coefficient.  Heat flow happens as a 
consequence of changes of conserved charges other than the
stress-energy tensor, but we do not consider this possibility here and
focus on the effects of the temperature. Bulk viscous effects, which
can be interpreted as a  consequence of the decay of particles within
the fluid, have been considered for warm inflation in
\cite{delCampo:2010by}, where 
they studied either a constant bulk viscous pressure or one proportional
to the radiation energy density.  The bulk
pressure appears at both the background and the perturbation level,
and being a negative pressure, it will favor warm inflation. For the
amplitude of the spectrum, for the phenomenological model considered
in \cite{delCampo:2010by} they found that it could induce a variation in
the amplitude of the order of 4\%. On the other hand, the shear
viscosity is related to changes in momentum of the  particles of the
fluid, 
and appears only at the level of the perturbations.     
Shear and bulk viscosity coefficients due to light field have been extensively
compute in the literature \cite{jeon,shear}, leading to power-law
dependences on $T$ for these coefficients. In addition, the bulk
viscous coefficient typically ends being much smaller than the shear
viscosity. Therefore, we will concentrate on the effect of the shear
viscosity on the spectrum, and do not consider those of the bulk
viscosity. The shear viscosity, being related to dissipation, appears
in the radiation fluid equation as a friction term which tends to damp
the growth of the fluctuations \cite{mossgraham}. Eventually, the
damping effect will dominate 
over  the enhancement induced by the dissipative source term. The aim
will be therefore to quantify, in a model independent way, this effect
on the spectrum, and when it will render the system field-radiation
effectively decoupled.     

The outline of the paper is as follow. In section II we review the
basic of warm inflation at the background level. In section III we
present the equations for the fluctuations of the coupled
field-radiation system, when the radiation is taken as an imperfect
fluid. The numerical solutions for the fluctuations are presented in
section IV, with and without the shear viscosity. We  also study
the inflationary model dependence of the results on the
spectrum by considering two generic model of inflation: a chaotic
model with general power $p$, and a standard quadratic hybrid model. 
In section V we summarize our findings.


\section{Warm inflation: background equations}
\label{sect2}

In any particle physics realization of the inflationary framework, the
inflaton is not an isolated part of the model but it interacts with
other fields.  These interactions  may 
lead to the dissipation of the inflaton energy into other degrees of
freedom, such that a small percent of the inflaton vacuum energy is
transferred into other kinds of energy. In the two-stage mechanism for
warm inflation, dissipation leads to particle production of light
degrees of freedom. When those relativistic particles thermalize fast
enough, say in less than a Hubble time in an expanding universe, we
can model their contribution as that of radiation:
\be
\rho_r \simeq \frac{\pi^2}{30}g_* T^4 \,,
\ee
where $T$ is the temperature of the thermal bath, and $g_*$ the
effective number of light degrees of freedom\footnote{If not otherwise
specified, we will take $g_* = 228.75$, the effective no. of degrees of
freedom for the Minimal Supersymmetric Standard Model, when presenting
numerical results.}.  

The dissipative term appears as an
extra friction term in the evolution equation for the inflaton field $\phi$,
\be
\ddot \phi + ( 3 H + \Upsilon ) \dot \phi + V_\phi =0
\,,\label{eominf}
\ee
$\Upsilon$ being the dissipative coefficient, $H=\dot a/a$ is the
Hubble rate of expansion, and
$a$ the scale factor of the Friedmann-Robertson-Walker background
metric:
\be
ds^2 = - dt^2 + a(t)^2 \delta_{ij}dx^i dx^j \,.
\ee 
Eq. (\ref{eominf})
is equivalent to the evolution equation for the inflaton energy
density $\rho_\phi$:
\be
\dot \rho_\phi + 3 H ( \rho_\phi + p_\phi) = - \Upsilon ( \rho_\phi +
p_\phi) \,,
\label{rhoinf}
\ee
with pressure $p_\phi = \dot \phi^2/2 - V(\phi)$, and $\rho_\phi
+ p_\phi= \dot \phi^2$. Energy conservation then demands that the
energy lost of the inflaton field must be gained by the radiation fluid
$\rho_r$, with the  RHS of Eq. (\ref{rhoinf}) acting as
the source term:
\be
\dot \rho_r+ 3 H ( \rho_r + p_r) = \Upsilon ( \rho_\phi +
p_\phi) \,. \label{eomrad}
\ee
In warm inflation, radiation  is not
redshifted away during inflation,
because it is continuously fed by the inflaton 
through the dissipation \cite{wi}. 
Inflation happens when $\rho_R \ll \rho_\phi$, but even if small when
compared to the inflaton energy density it can be larger than the
expansion rate with $\rho_R^{1/4} > H$. Assuming thermalization, this
translates roughly  into $T > H$. Otherwise, when $T < H$ (or
similarly when $\rho_R^{1/4} < H$), one just recovers the standard cold
inflation scenario, where dissipation can be neglected.

During warm inflation the motion of the inflaton field has to be
overdamped in order to have the accelerated expansion, but now this
can be achieved due to 
the extra friction term $\Upsilon$ instead of that of the Hubble
rate. And once $\phi$, $H$, and also $\Upsilon$, are in this slow-roll
regime, the
same will happen with the radiation energy density, the source term
now compensating for the Hubble dilution. In the slow-roll regime, the
equations of motion reduce to:
\bea
3 H ( 1 + Q ) \dot \phi &\simeq  -V_\phi    \,,\label{eominfsl} \\
4 \rho_R  &\simeq 3 Q\dot \phi^2\,, \label{eomradsl}
\end{align}
where we have introduced the dissipative ratio $Q=\Upsilon/(3 H)$. 
Notice that
 $Q$ is not necessarily constant. The
coefficient  $\Upsilon$ will depend on  $\phi$ and $T$, and 
therefore depending on the model the ratio $Q$ may increase or
decrease during inflation \cite{BasteroGil:2009ec}.
The slow-roll conditions are given by \cite{Moss:2008yb}: 
\bea
\epsilon &= \frac{m_P^2}{2} \left ( \frac{V_{\phi}} {V}
\right)^2 \frac{1}{1+Q}\ll 1\,, \label{eps}\\
\eta &= m_P^2 \left ( \frac{V_{\phi \phi}} {V}\right) \frac{1}{1+Q}
\ll 1 \,, \label{eta} \\
\beta_\Upsilon &= m_P^2 \left ( \frac{\Upsilon_\phi V_\phi }
     {\Upsilon V}\right) \frac{1}{1+Q}\ll 1 \,, \label{beta} \\
\delta &= \frac{T V_{T\phi}}{V_\phi} < 1  \label{delta}\,,
\end{align}
where the slow-roll parameter $\beta_\Upsilon$  takes into account the
variation of $\Upsilon$ with respect to $\phi$, and 
the last condition ensures that  thermal corrections to
the inflation potential are negligible. Similarly, taking also into
account the dependence on $T$ of $\Upsilon$, one has: 
\be
\left|\frac{d \ln \Upsilon}{d \ln T}\right| < 4 \,,
\ee
which reflects the fact that radiation has to be produced at a rate
larger than the redshift due to the expansion of the universe.   
The slow-roll regime ends when any of the above conditions
Eqs. (\ref{eps})-(\ref{delta}) is no longer satisfied, such that  either the
motion is no longer overdamped and slow-roll ends, or the radiation
becomes comparable to the inflaton energy density. Either way,
inflation will  end shortly  afterwards. 

For warm inflation the first requirement is to have $T>H$, but the
ratio $Q$ can be larger or smaller than unity. In the former case we
are in the strong dissipative regime (SDR), whereas the latter is
called weak dissipative regime (WDR). In the weak dissipative regime
the extra friction added by $\Upsilon$ is not enough to substantially
modify the background inflaton evolution, and it will resemble that
of cold inflation; still the thermal fluctuations of the
radiation energy density will modify the field fluctuations, and
affect the primordial spectrum of perturbations. 

The $T$ and $\phi$ dependent dissipative coefficient has been computed
in \cite{mossxiong}, using the near-equilibrium approximation
proposed in \cite{BGR}. The specific field theory models 
considered for the inflaton interactions leading to dissipation 
all follow from the two-stage mechanism \cite{BR1,br,br05}. 
In this mechanism,  
the inflaton field  $\phi$ 
is coupled to heavy catalyst fields  $\chi$, which decay into light
fields  $\sigma_i$. Consistency of the approximations then demands the
microphysical dynamics determining $\Upsilon$ to be faster than that
of the macroscopic motion of the background inflaton and the
expansion: 
\be
\Gamma_\chi > \left|\frac{\dot \phi}{\phi}\right|,\, H, 
\ee
where $\Gamma_\chi$ is the decay width of the heavy fields. 
In addition, the condition $T \gg H$ allows  to neglect the expansion of 
the universe  when computing $\Upsilon$. In the low $T$ regime, when 
 the mass of the catalyst field $m_\chi$ is larger than $T$, one has:
\be
\Upsilon(\phi,T) \propto \frac{T^3}{m_\chi^2} \propto
\frac{T^3}{\phi^2}\,, 
\ee
while in the high $T$ regime, where the thermal corrections to the
catalyst field mass start to be important,
\be
\Upsilon(\phi,T) \propto T \,.
\ee
And in the very high $T$ regime, the dissipative coefficient goes like
the inverse of $T$. These are the cases of study we are going to
consider in the next section when studying the fluctuations during
warm inflation. In general, we will parametrize the dissipative coefficient as:
\be
\Upsilon = C_\phi \frac{T^c}{\phi^{m}} \,,
\ee
with $c-m=1$. We will work with $c=$3,1,-1, and also $c=0$, the case
of no $T$ dependence for the dissipative coefficient.

\section{Fluctuations at linear order: Primordial spectrum}

During warm inflation we have a multicomponent
fluid, a mixture of a 
scalar inflaton field  $\Phi$ interacting with the radiation fluid.
Both components exchange energy and momentum through the dissipative
term $\Upsilon$. Dissipative effects also imply small departures from
equilibrium, and that the 
radiation fluid will not behave exactly like a perfect fluid during
inflation. In relativistic 
theory, these effect can be parametrized
in terms of a shear viscous tensor $\pi_{ab}$, an energy flux vector $q_a$
and a bulk viscous pressure $\pi_b$, in the stress-energy tensor for the
radiation fluid~\cite{weinberg,maartens},
\be 
T_{ab}^{(r)}= (\bar \rho_r + \bar p_r + \pi_b) u_a^{(r)}u_b^{(r)} + (\bar p_r +
\pi_b) g_{ab} + q_a^{(r)}u_b^{(r)}+q_b^{(r)}u_a^{(r)}+ \pi_{ab} \,,
\ee
where $\bar \rho_r$ is the energy density, $\bar p_r$ the adiabatic pressure,
$u_a^{(r)}$ the four velocity of the radiation fluid, $g_{ab}$ the
four-dimensional metric, and  $u_a^{(r)}
\pi^{ab}=0=g_{ab}\pi^{ab}$, $u_a^{(r)} q^{a}=0$. There would be heat 
flow for example in the presence of conserved charges in the system
other than the stress-energy tensor, but we do not consider such
possibility in this work, and then set $q_a=0$. The shear viscous
tensor vanishes in an homogeneous and isotropic background geometry,
but at linear order it is given by \cite{maartens}:
\be
\pi_{ab} \simeq -2 \zeta_s \sigma_{ab} \,, \label{shear}
\ee
where $\zeta_s$ is the shear viscosity coefficient and $\sigma_{ab}$
the shear of the radiation fluid:
\be
\sigma_{ab} = \nabla_{(a}u_{b)} + u_{(a} u^c\nabla_c u_{b)} -
\frac{h_{ab}}{3}\nabla^c u_c \,,
\ee
$\nabla_a$ being the covariant derivative of the metric $g_{ab}$. 
The bulk viscous pressure can be seen as a non-adiabatic pressure
contribution, already present at the background level. Nevertheless,
the contribution from the light fields is expected to be small with
respect to $p_r$.  Thus, we will also set $\pi_b=0$ and  focus on studying the
consequences of the shear viscosity on the primordial perturbations
during warm inflation.   

In order to study the system of perturbations at linear order, field,
radiation energy density  and radiation pressure 
are expanded around their background values in a
{}Friedman-Robertson-Walker metric: 
\begin{align}
\Phi(x,t) &= \phi(t) + \delta \phi(t,x) \,, \\
\bar \rho_r (x,t)&= \rho_r(t) + \delta \rho_r(t,x) \,, \\
\bar p_r(x,t) &= p_r(t) + \delta p_r(t,x) \,,
\end{align}
and similarly for the dissipative coefficient: $\bar \Upsilon(x,t)=
\Upsilon(t) + \delta \Upsilon(t,x)$.  The  perturbed FRW metric,
including only scalar perturbations,  is given by\footnote{Latin
  indexes $i,\,j,\,k,\ldots$ are used for the spatial components, and either
  $a,\,b,\,c,\ldots$ or Greek letters for space-time indexes. }:  
\be
ds^2= -(1+2 \alpha) dt^2 - 2 a \partial_i \beta dx^i dt 
     + a^2 [ \delta_{ij} (1 +2 \varphi) + 2 \partial_i \partial_j \gamma] dx^i
dx^j \,, \label{metric}
\ee
where $\varphi$ is the intrinsic curvature of a
constant time hypersurface. For later use, we introduce the combinations:
\bea
\chi & = a ( \beta + a \dot \gamma) \,, \\
\kappa&= 3 (H \alpha - \dot \varphi) + \partial_k \partial^k \chi \,, 
\end{align}
where $\chi$ is the shear and $-\kappa$ the
perturbed expansion scalar of the comoving frame. 
 
The evolution equations follow from the
conservation of the energy-momentum tensor: 
\be
\nabla^a T_{ab}^{(\alpha)}= Q_b^{(\alpha)} \,,\;\;\; \sum_\alpha
Q_b^{(\alpha)}=0\,,
\label{DTab}
\ee
where $Q_b$ is the four-vector source term  accounting for 
the exchange of energy and momentum:
\be
-Q_b^{(\phi)}=Q_b^{(r)}=\Upsilon u_{\phi}^a \nabla_a \Phi \nabla_b \Phi \,, 
\ee
$u_\phi^a$ is now the four-velocity of inflaton fluid:  
\be
u_\phi^a=-\frac{\nabla^a \Phi}{\sqrt{\rho_\phi + p_\phi}} \,,
\ee
and then:
\be
Q_b^{(r)}=\Upsilon (\bar \rho_\phi + \bar p_\phi)^{1/2} \nabla_b \Phi \,. 
\ee
The projection of the four-vector source term along the direction of the fluid 
gives the energy density source term, 
\be
Q^{(r)}=-u_{\phi}^a Q_a^{(\phi)}\,,
\ee
which at linear order is given by:
\bea 
Q^{(r)}&= Q_r + \delta Q_r \,, \\
Q_r & = \Upsilon \dot \phi^2 \,, \\
\delta Q_r &= \delta \Upsilon \dot \phi^2 + 2 \Upsilon \dot \phi \delta
\dot \phi - 2 \alpha \Upsilon \dot \phi^2 \,. \label{Qr}
\end{align}
The momentum source term $J_a$ is orthogonal to the fluid velocity:
\be
Q^{(\phi)}_a= Q^{(\phi)} u^{(\phi)}_a + J^{(\phi)}_a\,, \;\;\;\;
u^{(\phi)\,a} J_a^{(\phi)}=0 \,,
\ee
and vanishes in the FRW geometry; at linear order it reads:
\bea
J_i^{(r)} &= \partial_i {\bf J}_r \,, \\ 
{\bf J}_r& = - \Upsilon \dot \phi \delta \phi \,.\label{Jr}
\end{align}
To complete the specification of the source, we need $\delta
\Upsilon$, which for a general temperature $T$ and field $\phi$
dependent dissipative coefficient,   
$\Upsilon =  C_\phi T^c/\phi^m$, with $c-m=1$, is given by:  
\be
\delta \Upsilon =  \Upsilon \left(c \frac{\delta T}{T} - m
\frac{\delta \phi}{\phi} \right) 
\,. \label{dupsilon} 
\ee
Although dissipation implies departures from thermal equilibrium in
the radiation fluid, the system has to be close-to-equilibrium for the
calculation of the dissipative coefficient to hold, therefore $p_r \simeq
\rho_r/3$, $\rho_r \propto T^4$ and 
\be
4 \frac{\delta T}{T} \simeq \frac{\delta \rho_r}{ \rho_r}\,.
\ee

{}Finally, the evolution equations for the radiation fluctuations are obtained
expanding at linear order Eq. (\ref{DTab}) \cite{kodama,
  hwang,hwangnoh,malik}. Working in momentum space, defining the Fourier
transform with respect to the comoving coordinates, the equation of
motion for the fluctuations with comoving wavenumber $k$ are given
by\footnote{For simplicity, we keep the same notation  for the
  fluctuations $\delta f({\bf x},t)$ and their Fourier transform 
$\delta f({\bf k},t)$.}:    
\bea
\delta \dot \rho_r + 3 H (\delta \rho_r + \delta
p_r) &= -3 (\rho_r + p_r) \dot \varphi + \frac{k^2}{a^2} \left[
\Psi_r +(\rho_r + p_r) \chi\right] + \delta Q_r +
Q_r \alpha \,, \label{energyalpha}\\
\dot \Psi_r + 3 H \Psi_r &=- (\rho_r +
p_r)\alpha  - \delta p_r+
\frac{2 k^2}{3 a^2} \sigma_r + {\bf J}_r
\,, \label{momentumalpha} 
\end{align}
where a ``dot'' denotes the derivative with respect to the metric time
``$t$'', $\Psi_r$ is the radiation momentum perturbation, $T^{0\,(r)}_j=-
\partial_j \Psi_r/a$, and $\sigma_r$ the shear viscous pressure at
linear order:
\be
\sigma_r \simeq - 2 \zeta_s \left(\frac{\Psi_r}{\rho_r + p_r} + \chi\right)
\,. 
\ee

On the other hand, field fluctuations are described by a stochastic
system whose evolution is determined by a Langevin equation
\cite{calzetta,mossgraham}:

\bea
\delta \ddot \phi + (3 H + \Upsilon) \delta \dot \phi +
\left(\frac{k^2}{a^2} + V_{\phi\phi}\right) \delta \phi &= 
\left[2(\Upsilon + H) T\right]^{1/2}
a^{-3/2}\xi_k  \nonumber - \delta \Upsilon \dot \phi \\
& + \dot \phi ( \kappa + \dot \alpha) + (2 \ddot
\phi + 3 H \dot \phi) \alpha 
-\Upsilon ( \delta \dot \phi - \alpha \dot \phi) \label{field}\,,
\end{align}
where the stochastic source $\xi_k$ can be approximated by a
localized gaussian distribution with correlation function:
\be
\langle \xi(t,x) \xi(t^\prime,x^\prime)\rangle = \delta(t -t^\prime)
\delta^{(3)}(x-x^\prime) \,. 
\ee

So far, the equations for the perturbations at linear
order are written in a ``gauge ready'' form, without specifying any
particular gauge, but the equations can also be written in
terms of gauge invariant (GI) variables. For any scalar quantity $f$, at
linear order one can define a gauge invariant perturbation
\cite{kodama,hwang}: 
\be
\delta f^{GI} = \delta f - \frac{\dot f}{H} \varphi \,, 
\ee
while similarly the gauge invariant momentum perturbation reads:
\be
\Psi^{GI}= \Psi - \frac{\rho + p}{H} \varphi \,,
\ee
and for the metric perturbations one has the gauge invariant combinations:
\bea
{\cal A} &= \alpha - \left[ \frac{\varphi}{H} \right]^\cdot \,,\\
\Phi & = \varphi - H \chi \,.
\end{align}
The evolution equations then read:
\bea
\delta \ddot \phi^{GI} + 3 H \delta \dot \phi^{GI} + 
\left(\frac{k^2}{a^2} + V_{\phi \phi}\right) \delta \phi^{GI} &= 
\left[2 (\Upsilon+H) T\right]^{1/2}a^{-3/2}  \xi 
-\dot \phi \delta \Upsilon^{GI}  -\Upsilon \delta \dot \phi^{GI}
\nonumber  \\  
&+\Upsilon \dot \phi {\cal A} + \dot \phi \dot {\cal A}+ 2
(\ddot \phi + 3 H \dot \phi) {\cal A} -\frac{k^2}{a^2} \dot \phi
\frac{\Phi}{H}, \label{fieldGI} \\
\delta \dot \rho_r^{GI} + 3 H (1 + w_r) \delta \rho_r^{GI}  &=
\frac{k^2}{a^2} \Psi_r^{GI}+ \delta Q_r^{GI} + \dot \rho_r {\cal A} 
\,, \label{energyrGI}\\
\dot \Psi_r^{GI} + 3 H \left( 1 + \frac{k^2}{a^2 H^2}\bar \zeta_s\right) 
\Psi_r^{GI} &=- w_r \delta \rho_r^{GI} -
\Upsilon \dot \phi \delta \phi^{GI} - (\rho_r + p_r) {\cal A}
-\frac{3 k^2}{ a^2} (\rho_r+ p_r) \bar \zeta_s \frac{\Phi}{H} 
\label{momentumrGI} \,,
\end{align}
where in Eq. (\ref{momentumrGI}) we have defined: 
\be
\bar \zeta_s= \frac{4}{9} \frac{\zeta_s H}{\rho_r + p_r} \label{zetas}\,.
\ee 
{}Finally, from the Einstein equations at linear order, the gauge invariant
metric perturbations are given by \cite{kodama, hwang}:
\bea
{\cal A}&= -\frac{\dot H}{H^2} {\cal R} \,, \label{metricA}\\
\frac{k^2}{a^2 H^2}\Phi&= 3 {\cal A} + \frac{3}{2} \frac{\delta
  \rho_T^{GI}}{\rho_T}  \label{metricPhi}\,,
\end{align}
where ${\cal R}$ is the total comoving curvature perturbation,
\begin{eqnarray}
{\cal R}&=& \varphi - \frac{H}{\rho_T + p_T} \Psi_T 
\nonumber \\
&=& -\frac{H}{\rho_T
  +p_T} \Psi_T^{GI} \,.
\end{eqnarray}
{}For a multicomponent fluid, the total momentum perturbation is given
by the sum of the individual components, and ${\cal R}$ can be written
as the weighted sum of the individual contributions:
\bea
{\cal R} &= 
 \sum_{\alpha=\phi,\,r} \frac{h_\alpha}{h_T} {\cal R}_{\alpha} \,, \label{RT}\\
{\cal R}_\alpha & = -\frac{H}{h_\alpha} \Psi_\alpha^{GI}\,,
\end{align}
where we have defined $h_\alpha= \rho_\alpha + p_\alpha$. 
In particular, for a scalar field $\Psi_\phi^{GI}= - \dot \phi \delta
\phi^{GI}$, $h_\phi = \dot \phi^2$,  and
\be
{\cal R}_{\phi}= \frac{H}{\dot \phi} \delta \phi^{GI} \,, \label{Rphi}
\ee
for which the power spectrum would be the form for single
field cold inflation. For warm inflation, we shall use the total
comoving curvature perturbation to evaluate the primordial spectrum,  
\be
P_{\cal R}(k)= \frac{k^3}{2 \pi^2} \langle |{\cal R}_k|^2 \rangle\,, \label{PR}
\ee
where ``$\langle \cdots \rangle$'' means average over different
realizations of the noise term in Eq. (\ref{field}). In all numerical
results shown in this work we have performed averages over 1000 runs. 
This was found to be more than enough to get convergent numerical results.".

The largest observable scale in the CMB corresponds to a comoving
scale $k$ crossing the horizon $N_e$ e-folds before the end of
inflation, denoted by $k=a_* H_*$. The value of $N_e$ can vary in
general from 40 to 70 depending on the inflationary model and details on the
subsequent reheating process \cite{reheating}. Although we shall
consider different inflationary potentials, we will not consider the
details of reheating period, and simply fix the horizon crossing at
$N_e\simeq 60$. As we will see, soon after horizon crossing, the amplitude of
the individual curvature perturbations ${\cal R}_\phi$ and ${\cal
  R}_r$  freezes out, and so does that of the total
curvature. This allows to compute the primordial spectrum by
evaluating Eq. (\ref{PR}) at horizon crossing, mainly when getting
analytical approximations. Nevertheless, when
showing numerical results we shall integrate Eqs. (\ref{fieldGI})-
(\ref{momentumrGI}) and evaluate the amplitude of
the spectrum say 10 e-folds after horizon crossing.

\section{Equations in the zero-order slow-roll approximation and beyond}

In order to gain some insight on the evolution of the perturbations,
we follow \cite{mossgraham} and first study the equations for the
perturbations at zero-order in the slow-roll parameters. That is, 
expanding the background variables around their slow-roll values and
neglecting all terms in the equations proportional to slow-roll
parameters. This eliminates in the equations for the fluctuations the
dependence on the inflationary potential, and the details on the
evolution of the dissipative coefficient. For example, the metric
perturbation  ${\cal A}$ given in 
Eq. (\ref{metricA}) is proportional to the slow-roll coefficient
$\epsilon$, Eq. (\ref{eps}),    
$\epsilon= -\dot H/H^2$,
and can be neglected; 
and the same for the terms proportional to $H \dot \phi$, 
{\it i.e.}, those
proportional to $\Phi$, and $\dot \phi/H\phi$. Defining the
dimensionless variables:  
\bea
y_k &= \frac{k^{3/2}\delta \phi^{GI}} {\left[2 (\Upsilon + H) T\right]^{1/2}} \,,\\
w_k &= \frac{k^{3/2}\delta \rho_r^{GI}}{ \left[2 (\Upsilon + H) T\right]^{1/2} 
(\Upsilon \dot \phi)} \,, \\
u_k &= \frac{k^{3/2}\Psi_r^{GI}}{ \left[2 (\Upsilon + H) T\right]^{1/2} \dot \phi} \,,
\end{align}
and using the slow-roll background equations (\ref{eominfsl}),
(\ref{eomradsl}), 
we have the system:
\begin{align}
\ddot y_k + 3 H ( 1 +Q) \dot y_k + H^2 \left[ z^2 + 3 \eta (1+Q) - 3m Q
\frac{\dot \phi}{H\phi} 
\right] y_k 
&= \left(\frac{k}{a}\right)^{3/2}\xi_k- 3 Q c H^2 w_k \,, \label{EOMyk} \\
\dot w_k + H (4-c) w_k &= \frac{H}{3 Q}z^2 u_k + 2 \dot y_k 
\,, \label{EOMwk}\\ 
\dot u_k + 3 H(1 + z^2 \bar \zeta_s) u_k &= -3 Q H \left(\frac{w_k}{3}
+y_k\right) 
\,, \label{EOMuk}  
\end{align}
where $z=k/(aH)$. Combining Eqs. (\ref{EOMwk}) and (\ref{EOMuk}) into
a second order differential equation, we have:
\bea
\ddot w_k + H ( 9 -c + 3 z^2 \bar \zeta_s ) \dot w_k + H^2 \left[ 20 - 5 c +
6 Q c + \frac{z^2}{3} + 3 z^2 (4-c) \bar \zeta_s\right] w_k 
&=
2 \left(\frac{k}{a}\right)^{3/2}\xi \nonumber \\
& 
\!\!\!\!\!\!\!\!\!\!\!\!\!\!\!\!\!\!
\!\!\!\!\!\!\!\!\!\!\!\!\!\!\!\!\!\!
\!\!\!\!\!\!\!\!\!\!\!\!\!\!\!\!\!\!
\!\!\!\!\!\!\!\!\!\!\!\!\!\!\!\!\!\!
\!\!\!\!\!\!\!\!\!\!\!\!\!\!\!\!\!\!
\!\!\!\!\!\!\!\!\!\!\!\!\!\!\!\!\!\!
+ H ( 4 - 6 Q + 6 z^2 \bar \zeta_s) \dot y_k 
- H^2 \left[3 z^2 + 6 \eta (1+Q)- 6 Q m \frac{\dot \phi}{H \phi}\right]
y_k  \label{fullwk}\,.  
\end{align}

\begin{figure}
\vspace{0.5cm}
 \includegraphics[width=0.55\textwidth,
  angle=0]{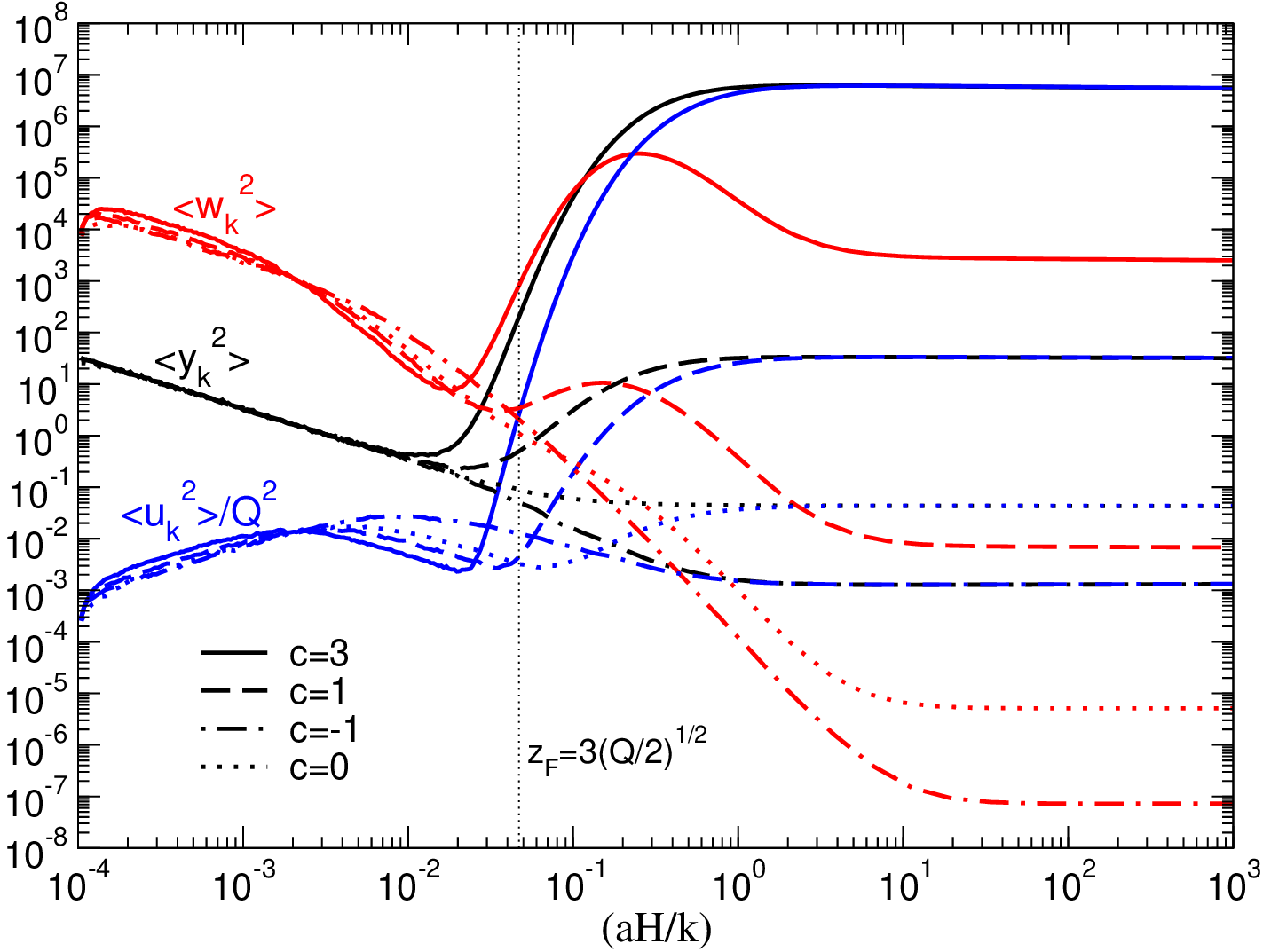}
\vspace{0.25cm}
\caption{\label{plot1a} Evolution of the power
  spectrum of the field 
  $\langle y_k^2\rangle$, radiation energy density $\langle
  w_k^2\rangle$, and radiation momentum $\langle u_k^2\rangle/Q^2$,
  for $Q=100$, and wavenumber 
  $k=10^4 H$. The vertical thin dotted line sets the value of the freeze
  out scale $k_F$ in warm inflation.     
  The results are shown for different power dependence on
  $T$ of the dissipative coefficient: 
  $c=3$ (solid lines), $c=1$ (dashed lines), $c=-1$ (dash-dotted
  lines), and $c=0$ (dotted lines).
} 
\end{figure}

\begin{figure}
\vspace{0.5cm}
\includegraphics[width=0.55\textwidth, angle=0]{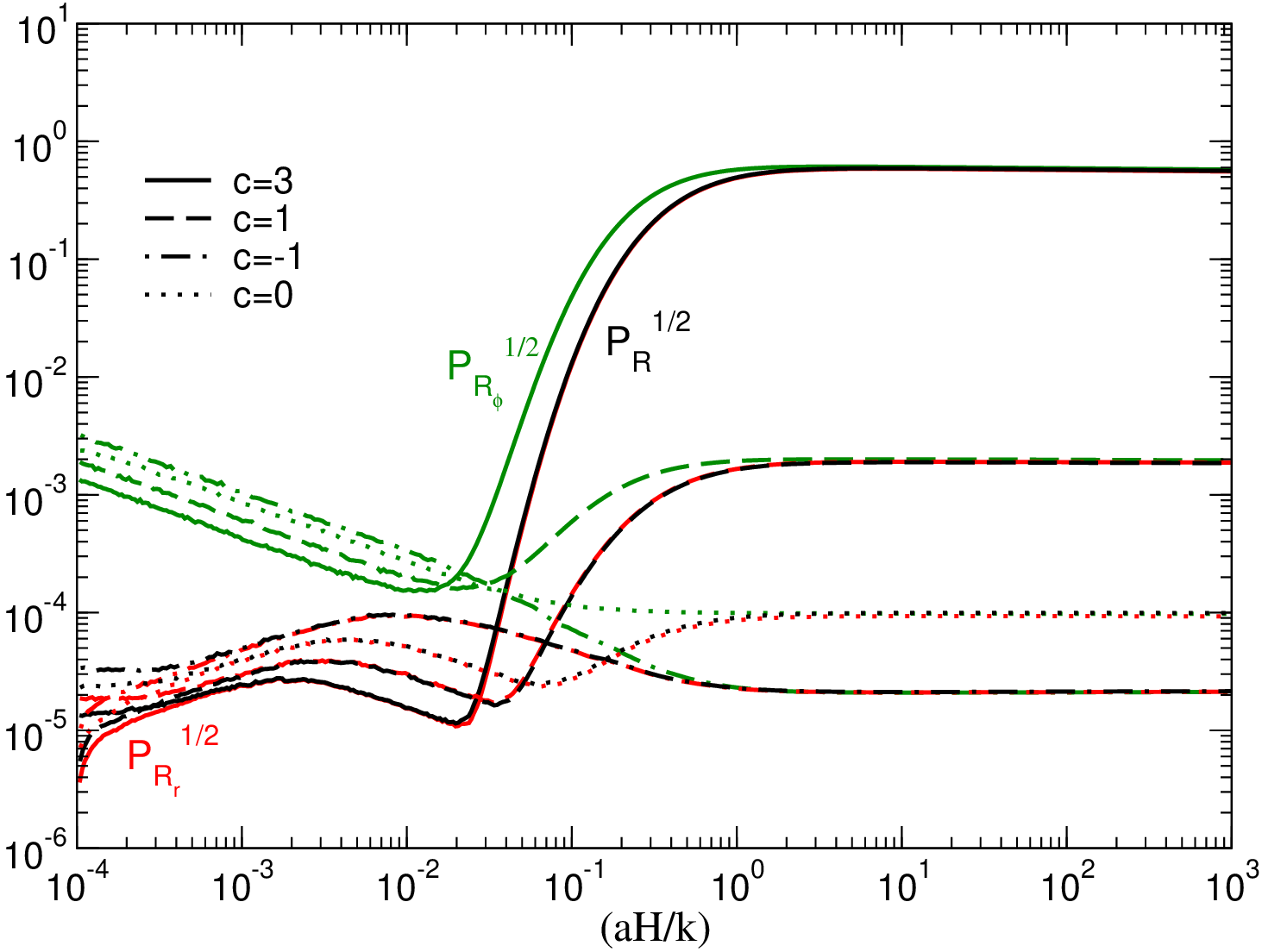}
\vspace{0.25cm}
\caption{\label{plot1b} Evolution  of the total curvature perturbation
  spectrum $P_{\cal R}^{1/2}$ (black lines), the 
  radiation $P_{{\cal R}_r}^{1/2}$ (red lines), and the field
  $P_{{\cal R}_\phi}^{1/2}$ (green lines) curvature perturbation spectrum.
  The results are shown for different power dependence on
  $T$ of the dissipative coefficient: 
  $c=3$ (solid lines), $c=1$ (dashed lines), $c=-1$ (dash-dotted
  lines), and $c=0$ (dotted lines).
} 
\end{figure}

Before including shear effects, we study the evolution of the
perturbations setting $\bar \zeta_s=0$. 
The evolution of the power spectrum of the field $\langle y_k^2
\rangle$, the radiation energy  density $\langle w_k^2 \rangle$, and
the radiation momentum $\langle u_k^2 \rangle$, are shown 
in {} Fig. \ref{plot1a} as a function of $z^{-1}=aH/k$. We have
taken $Q=100$,  and started the
integration at $z_i=10^4$. As mentioned above, quantities as $\langle
y_k^2 \rangle$ denotes the average over 1000 realizations of the
gaussian noise term,  and by $y_k^2$, $w_k^2$, $u_k^2$ we mean the
modulus of the complex variable. We have set the initial conditions
for field fluctuations in the vacuum,  while $w_k$ and $u_k$ initially
vanish, for simplicity. Starting the evolution early enough before horizon
crossing, the system is always controlled by the stochastic source
term, and the dependence on the initial conditions is quickly erased.

We have considered different powers of $T$ for the dissipative
coefficient $\Upsilon$, $c=3,\,1\,,-1$, and included the
case of a constant or 
field dependent $\Upsilon$ ( $c=0$, dotted lines) for comparison. 
The radiation fluctuation $w_k$ acts as a source term for the field,
but at early times $z^{-1} \ll 1$, for subhorizon perturbations, the
field evolution is dominated by the stochastic source term, and both
radiation and field fluctuations evolve like in the case $c=0$. 
In the latter
case, the freeze out of the perturbations takes place before horizon
crossing, due to the extra friction term in Eq. (\ref{EOMyk}), at
around $k_F/(aH)\simeq 3\sqrt{Q/2}$ (vertical thin dotted line)
\cite{warmdeltap,warmpert, mossgraham}, and
soon after field and radiation spectrum level off. The field spectrum
for a $T$ independent $\Upsilon$  can be computed analytically and is
given by \cite{warmdeltap}:    
\be
\langle y_k^2 \rangle_0 \simeq \frac{\sqrt{3\pi}}{4} 
\frac{\sqrt{1+Q}}{(1+3 Q)}\,,  \label{yy0}
\ee
where the subindex ``0'' denotes the value for $c=0$. On the other
hand, when $c > 0$, field and radiation fluctuations get
effectively coupled before freeze out at around $z_c^2 \approx 18 Q c$,
and both start growing at similar
rates. Numerically, we get: 
\be
\langle y_k^2 \rangle \approx \langle y_k^2 \rangle_{z_c} \left( \frac{z_c}{z}
\right)^{5 c} \,. \label{yyz} 
\ee
The field spectrum when the fluctuations are still subhorizon and $z >
z_c$, which is independent of the radiation fluctuation,  can be
found in \cite{mossgraham}: 
\be
\langle y_k^2 \rangle_{z_c} \approx \frac{3c}{z_c} \,, \label{yyzc}
\ee
and therefore, at horizon crossing: 
\be
\langle y_k^2 \rangle_* \propto z_c^{5c-1} \propto Q^{(5c-1)/2} \,. \label{yystar}
\ee
When $c< 0$, the effect is the opposite, and effectively the freeze
out is delayed until practically horizon crossing, which makes the
amplitude of the field spectrum smaller than in the $c=0$ case. 

{}Fig. \ref{plot1b} shows the evolution
of the comoving curvature power spectrum, $P_{\cal R}^{1/2}$, given by
the sum of the radiation and the field contributions as in
Eq. (\ref{RT}), for different values of $c$. Also shown are 
the power spectra of
the radiation, $P_{{\cal R}_r}^{1/2}$, 
and the field $P_{{\cal R}_\phi}^{1/2}$. After  
horizon crossing they all converge to the same amplitude. During
slow-roll one has that $h_r = \rho_r +p_r \simeq Q h_\phi$, and from
Eq. (\ref{EOMuk}) when $z\ll 1$ the radiation momentum
becomes proportional to the field fluctuation:   
\be    
\Psi_r \simeq Q \dot \phi \delta \phi \,,
\ee
and therefore: 
\be
P_{{\cal R}_r} = \frac{H}{h_r} P_{\Psi_r} \simeq P_{{\cal R}_\phi} \,.
\ee
Owing to the fact that $h_\phi \ll h_r \simeq h_T$, the main
contribution to the primordial spectrum in Eq. (\ref{PR}) comes from
the radiation, before and after horizon crossing. The
primordial spectrum is always dominated by the thermal fluctuations. 
But after horizon crossing one simply has:
\be
P_{\cal R} \simeq P_{{\cal R}_r}\simeq P_{{\cal R}_\phi} \,.
\ee  
 Therefore, the amplitude of the primordial spectrum can be written as
 usual in terms of that of the inflaton field:
\be
P_{\cal R} \simeq \left( \frac{H}{\dot \phi} \right)^2 \frac{(H +
  \Upsilon)T}{\pi^2} \langle y_k^2 \rangle_*   \label{PRyy}\,,
\ee
evaluated at horizon crossing.

\begin{figure}
\vspace{0.55cm}
\includegraphics[width=0.55\textwidth,angle=0]{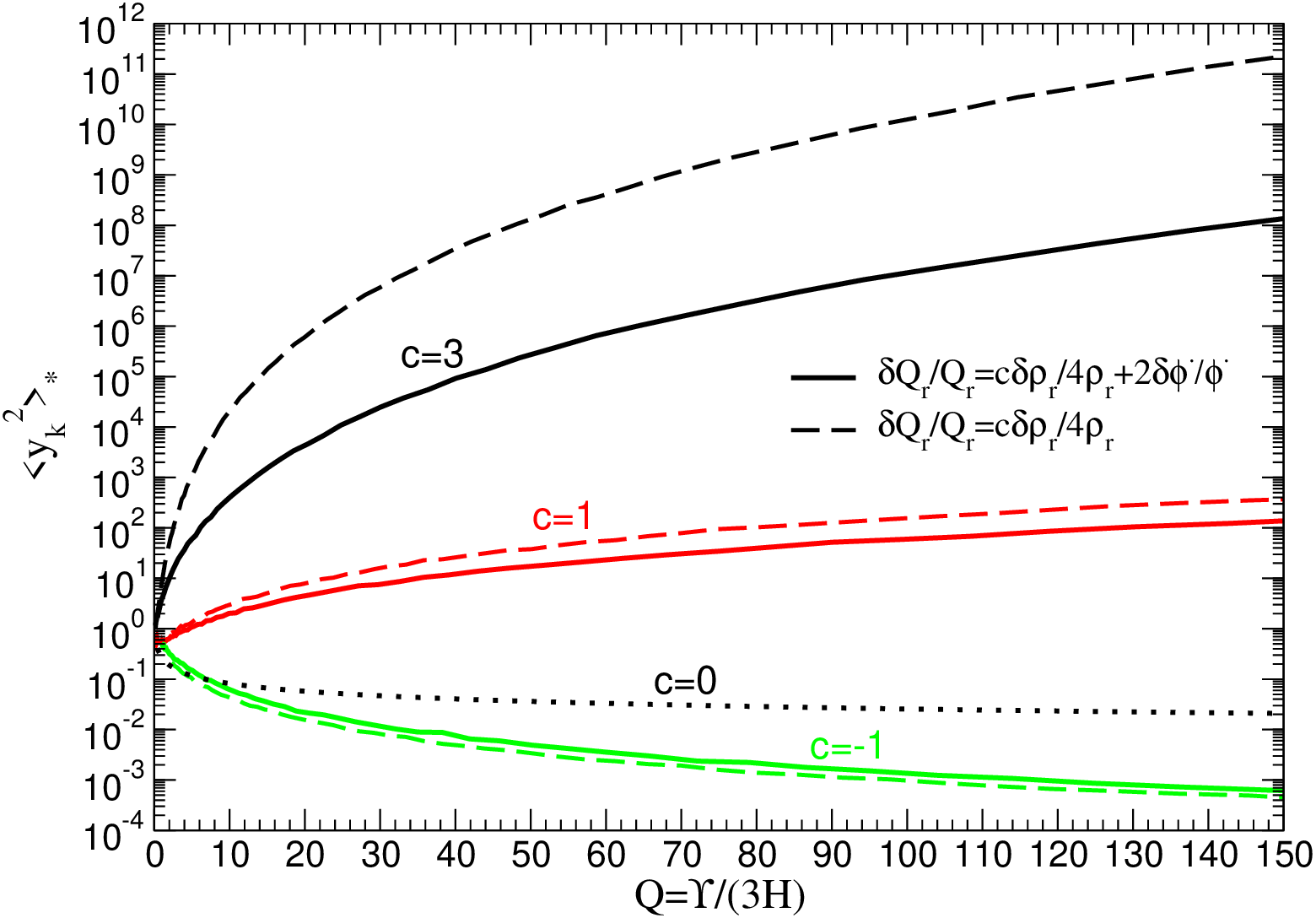}
\vspace{0.25cm}
\caption{\label{plot2} The spectrum of the field at horizon crossing 
  $\langle y_k^2\rangle_*$ as a function of the
  dissipative parameter $Q$, at zero order in the slow-roll
  parameters, for different values of $c$, and no shear $\bar
  \zeta_s=0$. Solid lines are obtained integrating
  Eqs. (\ref{EOMyk})-(\ref{EOMuk}), while dashed lines were obtained
  with the approximation used in \cite{mossgraham}.     
} 
\end{figure}

In {}Fig. \ref{plot2} we have compared the power spectrum of the
field, $\langle y_k^2 \rangle$ at horizon crossing, as a function of the
dissipative parameter $Q$ for different values of $c$. The equation
for the fluctuations has been integrated keeping the background values
constant (solid lines), Eqs. (\ref{EOMyk})-(\ref{EOMuk}).  The larger
the power $c>0$, the larger the enhancement with $Q$, as the
fluctuations get coupled earlier. For $c <0$, as mentioned 
before, we have the opposite effect, and the spectrum diminished with
respect to the case $c=0$. For positive $c$, the curves can be fitted by
a function:
\be
\langle y_k^2 \rangle_* |_{c>0}\simeq \langle y_k^2 \rangle_0 
( A_c Q^\alpha + B_c Q^\beta) \,.  \label{yyapprox}
\ee
but when $c=-1$, we have found that the curve can be best fitted by: 
\be
\langle y_k^2 \rangle_*|_{c=-1} \simeq \frac{ 1+ A_{-1} Q^\alpha}{ 1+ B_{-1}
  Q^\beta} \,. \label{yyapproxcn1}
\ee
The coefficients are given in Table \ref{table1}. 
For $c=3,\,1$, the approximation works well for $Q>50$, while for
$c=-1$ it is valid for any $Q$.

\begin{table}
\begin{tabular}{|c|c|c|c|c|}   
 \hline
$c$ & $\alpha$ &  $\beta$ & $A_c$ & $B_c$ \\
\hline
~~~3~~~ & ~~~7.5~~~ & ~~~7.0~~~ & ~~~$1.9\times 10^{-8}$~~~ & ~~~$3.4 \times 10^{-6}$~~~ \\
~~~1~~~ & ~~~2.5~~~ & ~~~2.0~~~ & ~~~$2.8 \times 10^{-2}$~~~ & ~~~$6.8 \times 10^{-5}$~~~ \\
~~~-1~~~ & ~~~0.2~~~ & ~~~1.4~~~ & ~~~0.78~~~ & ~~~0.088~~~ \\
\hline
\end{tabular} 
\caption{\label{table1} Coefficients for the numerical fit of the
  spectrum, Eq. (\ref{yyapprox}) and Eq. (\ref{yyapproxcn1}).}
\end{table}

We have also included in {}Fig. \ref{plot2} the spectrum of the field
obtained with the approximation used in \cite{mossgraham} (dashed
lines) for comparison. We have confirmed the main results about the
power spectrum obtained in \cite{mossgraham}; {\it i.e.}, that for $c >0$,
the amplitude of the primordial spectrum in warm inflation is enhanced
by a factor $\simeq Q^{\alpha -1/2}$. However, while in
\cite{mossgraham} they found $\alpha = 3c$, we have a smaller power
$\alpha=5c/2$, which for $c=3$ can   
mean a difference of two or 3 orders of magnitude in the amplitude of
the spectrum for $Q\simeq 50 - 100$. The differences can be traced
back to how the radiation source term is treated. In \cite{mossgraham}, it
was approximated by:
\be
\delta Q_r \simeq Q_r \frac{\delta \Upsilon}{\Upsilon} \simeq c Q_r
\frac{\delta \rho_r}{4\rho_r} \,, \label{Qrian} 
\ee 
while we have kept the dependence on the field:  
\be
\delta Q_r \simeq Q_r \left( c \frac{\delta \rho_r}{4\rho_r} + 2
\frac{\delta \dot \phi}{\dot \phi}\right) \,. \label{Qrphi}  
\ee 
With no shear viscosity included, the second order differential
equation Eq. (\ref{fullwk}) reads:  
\bea
\ddot w_k + H ( 9 -c ) \dot w_k + H^2 \left( 20 - 5 c  \frac{z^2}{3}\right) w_k 
+  H^2 z^2 y_k  
&= 2 \left(\frac{k}{a}\right)^{3/2}\xi  \nonumber \\
& 
\!\!\!\!\!\!\!\!\!\!\!\!\!\!\!\!\!\!
\!\!\!\!\!\!\!\!\!\!\!\!\!\!\!\!\!\!
\!\!\!\!\!\!\!\!\!\!\!\!\!\!\!\!\!\!
\!\!\!\!\!\!\!\!\!\!\!\!\!\!\!\!\!\!
- H^2 6 Q c w_k +  H ( 4 - 6 Q ) \dot y_k 
- H^2 \left[2 z^2 + 6 \eta (1+Q) - 6 Q m \frac{\dot \phi}{H \phi}\right]
y_k  \label{fullwknoshear}\,,  
\end{align}
where we have written the equation such that on the RHS we have the terms
induced by the field dependence in Eq. (\ref{Qrphi}). Thus, setting
the RHS to zero one recovers the equation derived with the source term
as given in Eq. (\ref{Qrian}). While the term
proportional to the field perturbation $y_k$ acts as a source term on
the radiation that tends to enhance the fluctuation, the extra terms 
$6 H^2 Qc w_k$ and $H(4 -6Q) \dot y_k$ have the opposite effect, i.e.,
that of damping the growth. Although these contributions are not enough
to avoid the growth of the fluctuations, they have a sizable effect on
their power-law behavior, mainly when $c=3$, as seen in Fig. \ref{plot2}. 
When $c =-1$ the radiation fluctuations do not grow before horizon
crossing, so that the effect of the field dependent terms in
Eq.~(\ref{Qrphi}) is negligible. 

\begin{figure}
\vspace{0.55cm}
\includegraphics[width=0.55\textwidth,angle=0]{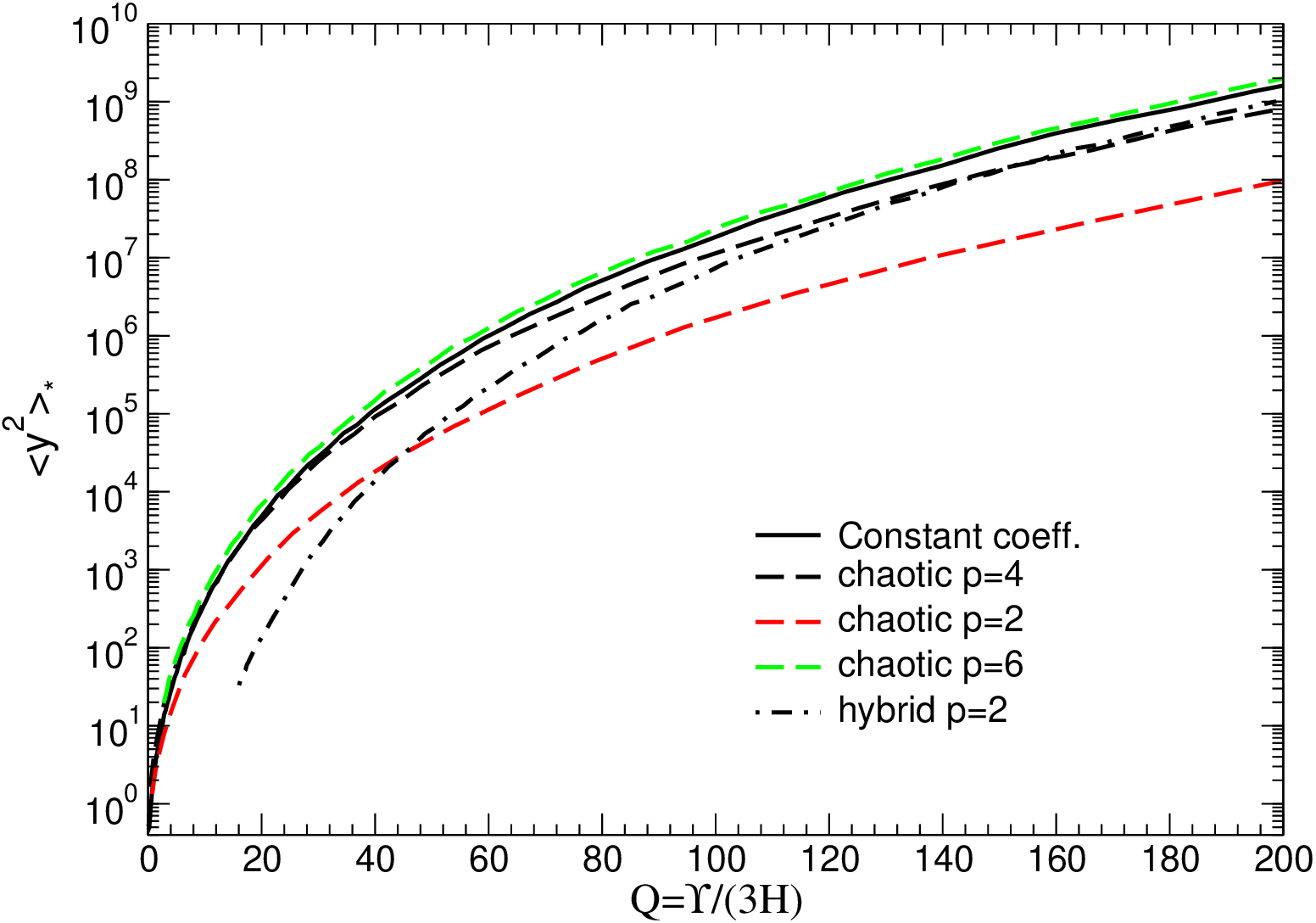}
\vspace{0.25cm}
\caption{\label{plot3} The spectrum of the field 
  $\langle y_k^2\rangle$ at horizon crossing as a function of the
  dissipative parameter $Q_*$, for different inflationary models, $c=3$
  and no shear $\bar \zeta_s=0$. The solid line is the result at zero
  order in slow-roll; dashed lines are for a chaotic model with
  $p=$6,4,2, from top to bottom; the dash-dotted line is a quadratic
  hybrid model. 
} 
\end{figure}

Neglecting the evolution of the background variables and working at
zero order in the slow roll parameters,  is a good approximation when
$Q$ is large enough, and the background parameters hardly vary during
the last 60 e-folds of inflation. Indeed the spectrum depends mainly
on the value of the parameters in a smaller interval, 5-6 e-folds
around horizon crossing, where one may expect the approximation of
keeping them constant to be a fairly good one. Still, this is a model
dependent question. This can be seen in {}Fig. \ref{plot3}, where we
show the field spectrum for some generic inflationary models, and
$c=3$, $\bar \zeta_s=0$. The value of  
$\langle y_k^2 \rangle_*$ has been obtained integrating
Eqs. (\ref{field})-(\ref{momentumrGI}), evaluating the amplitude of
the comoving curvature spectrum at $N_e=20$ efolds after horizon crossing, and
using Eq. (\ref{PRyy}). We have considered the inflationary models:
\bea
V& = \frac{V_0}{p} \left( \frac{\phi}{m_P}\right)^p\,, \;\;\; & {\rm
  (chaotic)} \,, \label{chaotic}\\ 
V& = V_0 \left[ 1 + \frac{\eta_\phi}{2} \left( \frac{\phi}{m_P}\right)^2
\right] \,, \;\;\; & {\rm (hybrid)}  \label{hybrid}\,. 
\end{align} 
{}For the chaotic model we have run different powers $p=$ 6, 4, 2, and
set $V_0= 10^{-14} m_P^4$, while for the hybrid $V_0= 10^{-8}m_P^4$
and $\eta_\phi=3$. For each model, the initial value of the inflation
field is chosen such that we can have $N_e \simeq 64$, and from the
background slow-roll equations one derives the initial values of the field
derivative, $\rho_r$ and $Q$. We have chosen $k= 100 H_i$, $H_i$ being
the initial value of the Hubble parameter. Therefore, horizon crossing
$k = a_* H_*$ takes places at around 60 e-folds before the end of
inflation.

In all the examples considered above the dissipative coefficient
increases during inflation \cite{BasteroGil:2009ec}, and the larger the
power in the potential, the slower the evolution of the background
values.  In the plot, the solid line is the result obtained with constant
 background variables. For a quartic chaotic model or larger power, the
approximation at zero order in the slow-roll works fine, while for a
quadratic chaotic it tends to overestimate the spectrum by at least an order of
magnitude for $Q \gtrsim 50 $. {}For the hybrid model, the model
dependence shows up when $Q \lesssim 100$.

\begin{figure}[t]
\vspace{0.5cm}
\includegraphics[width=0.55\textwidth,angle=0]{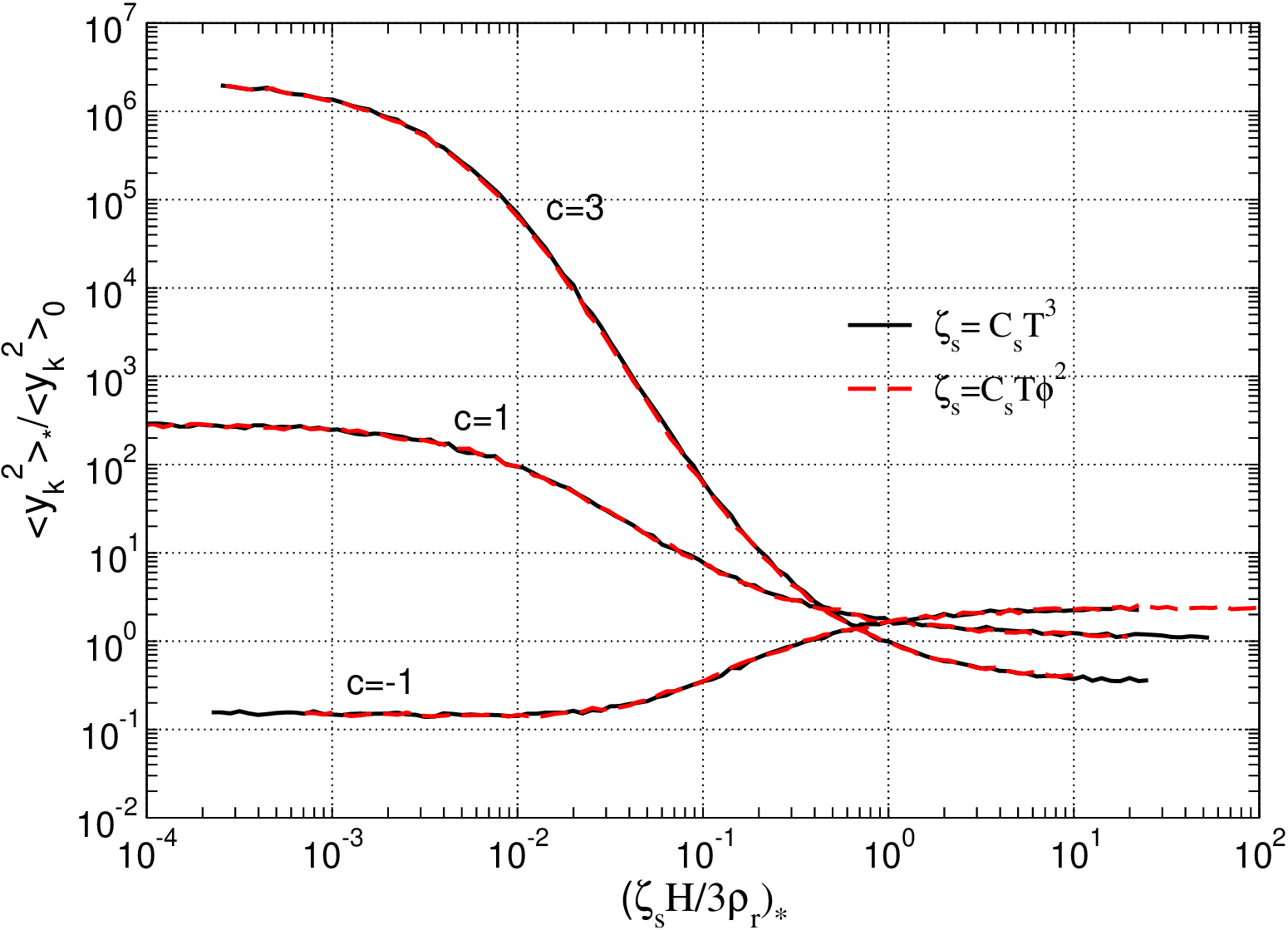}
\vspace{0.25cm}
\caption{\label{plot4} Field spectrum normalized by the value with
  $c=0$, as a function of the shear parameter $\bar \zeta_s$, for
  different values of $c$, and $Q=40$. From top to bottom, $c=3,\,1,\,-1$. For
  each curve, we have also considered two different $T$ dependence on
  the shear viscosity, as indicated in the legend, but both gives the
  same field spectrum. } 
\end{figure}

Shear effects will further damp the growth of the fluctuations.  
In Eq. (\ref{fullwk}) the shear acts as an extra friction when the
fluctuations are still subhorizon, suppressing the amplitude of the
radiation fluctuation before the 
radiation-field system becomes effectively coupled. Whenever the shear
is large enough, this suppression indeed can prevent altogether the
growth of the field perturbations, as the amplitude of the radiation
fluctuation is not enough to affect that of the field before horizon
crossing. This will happen when $\bar \zeta_s= \zeta_s H/(3 \rho_r)
\gtrsim 1$ at horizon crossing. During warm inflation, we have the
catalyst field 
coupled to the inflaton field, and to the light degrees of freedom
giving rise to the thermal bath. The shear viscosity for light fields
(light with respect to the temperature $T$ of the thermal bath)
typically behaves 
as $\zeta_s \propto T^3$ \cite{shear,Moore:2007ib}, although, depending on the
pattern of interactions, other powers could be
possible and cannot be excluded. Nevertheless, the damping  is fully
controlled by the value of the dimensionless parameter $\bar \zeta_s$ at
horizon crossing, and {\it is independent of  the functional form of the shear
with the temperature}, as can be seen in
{}Fig. \ref{plot4}. We have integrated the full EOM without
approximations Eqs. (\ref{field})-(\ref{momentumrGI}), for a quartic
chaotic model, Eq. (\ref{chaotic}) with $p=4$, and set the initial
background values such that $Q_*\simeq 40$. We show in the plot  
the dependence of the field spectrum with $\bar \zeta_s$ at horizon
crossing, for different values of $c$, and two 
examples of the shear viscosity: one proportional to $T^3$ (solid lines), and
another linear in $T$ (dashed lines), however the curves lay on top of each
other. We have checked that this is independent of the inflationary
model considered. We have normalized the field spectrum with the value obtained
when $c=0$. As the shear
increases, it does damp the radiation fluctuation enough for the
field spectrum to approach the $c=0$ value. Asymptotically, when $\bar
\zeta_s \gg 1$, for a linear dissipative coefficient with $T$ one
practically recovers the
$c=0$ case, for a cubic one the field spectrum is $\simeq 2 \langle
y_k^2\rangle_0/5$, while the inverse $T$ case is slightly above $\simeq
5 \langle y_k^2 \rangle_0/2$. In all cases, the field spectrum is well
fitted by a function:
\be
\log_{10} \frac{\langle y_k^2 \rangle}{\langle y_k^2\rangle_0 } \simeq 
A_s - B_s\left[1+ \tanh \left( \log_{10}\bar \zeta_s + \Delta_s\right)\right] 
\,, \label{fittingyys}
\ee
which interpolates between the result with no shear $\sim 10^{A_s}$ and 
 the  $c=0$ case, modulo a normalization constant $\sim
 10^{A_s- 2 B_s}$. As an example, the coefficients $A_s$, $B_s$ and
 $\Delta_s$ for the  potential  shown in Fig. \ref{plot4} with
 $Q_*=40$ are given in Table
   \ref{table2}.  

\begin{table}
\begin{tabular}{|c|c|c|c|}   
 \hline
$c$ & $A_s$ & $B_s$ & $\Delta_s$ \\
\hline
~~~3~~~ & ~~~6.35~~~ & ~~~3.4~~~ & ~~~1.36~~~ \\
~~~1~~~ & ~~~1.9~~~ & ~~~1.2~~~ & ~~~1.33~~~ \\
~~~-1~~~ & ~~~-0.95~~~ & ~~~-0.7~~~ & ~~~-0.66~~~ \\
\hline
\end{tabular} 
\caption{\label{table2} Coefficients for the numerical fit of the
  spectrum of the field with the shear viscosity,
  Eq. (\ref{fittingyys}).}
\end{table}

Finally, combining Eq. (\ref{fittingyys}) with Eq. (\ref{yyapprox}),
the field spectrum when $c>0$ reads:
\be
\langle y_k^2 \rangle_* \simeq \langle y_k^2 \rangle_0 F_Q[Q]^{F_s[\bar \zeta_s]} \,,
\ee
where: 
\bea
F_Q[Q] &\simeq \left( A_c Q^\alpha + B_c Q^\beta \right) \,, \\
F_s[\bar \zeta_s]& \simeq \frac{1}{2}\left[1- \tanh \left( \log_{10}\bar \zeta_s + 
\Delta_s\right)\right] \,. 
\end{align}
Therefore, in the strong dissipative regime when $Q>1$, the primordial
spectrum of the curvature perturbation can be written as that obtained
for a $T$ independent dissipative coefficient, times an enhancement
function $F_Q[Q]$ depending on the dissipative ratio $Q$, but
controlled by a function depending on the shear $F_s[\bar \zeta_s]$ :
\be
P_{\cal R} \simeq \left( \frac{H^2}{\dot \phi} \right)^2
\frac{\sqrt{3\pi}}{4 \pi^2}\left[\frac{(1 +
  Q)^{3/2}}{1+3Q}\right]  \left(\frac{T}{H}\right) \times F_Q[Q]^{F_s[\bar \zeta_s]} \,, 
\ee
and whenever $\bar \zeta_s > 1$ one recovers the amplitude of the
primordial spectrum obtained when $c=0$. The latter is of a magnitude
compatible with the observational value, for model parameters values
common in inflationary model building \cite{BasteroGil:2009ec,Zhang:2009ge}. For
example, without the enhancement, a quartic warm chaotic
model gives rise to the right level of perturbations with a coupling
constant $\lambda \simeq 10^{-13}-10^{-14}$. From the point of view of
model building, the enhancement produced
by the backreaction of the radiation fluctuations onto the fields, if
not avoided by shear effects, can be compensated by lowering the
height of the potential, i.e., by lowering couplings and
masses. However, it will also have an impact on the prediction for the
spectral index: for models with $Q$ increasing at the time of horizon
crossing it will render the spectrum too blue-tilted, and the other
way round, for $Q$ decreasing the spectrum may become too
red-tilted.  Shear viscosity damps the growth of the fluctuations, and
therefore will also affect the spectral index. 

Even though the results we have obtained are fairly model independent
and shown not to depend on the specific dependence on the temperature
of shear viscosity, it is useful to estimate the
magnitude of the ratio $\zeta_s H/(3 \rho_r)$ for typical warm
inflation models. In kinetic theory, the shear viscosity for
relativistic fluids can be expressed parametrically as proportional to
the mean free path of quasiparticles in the fluid. Considering as an
example a radiation fluid made of relativistic scalar particles
$\sigma$, with mass $m_\sigma/T \ll 1$ and self-interaction potential
$\lambda_\sigma \sigma^4/4!$, the mean free path of quasiparticles is
determined by the inverse of the thermal width, which is ${\cal
  O}(\lambda_\sigma^2)$, and the computed value for the shear
viscosity is \cite{jeon} $\zeta_s \simeq 3 \times 10^3
T^3/\lambda_\sigma^2$.  The condition $\zeta_s H/(3 \rho_r) \gtrsim 1$
can then be expressed, for example, as a condition on the magnitude of
the radiation bath self-interaction, $\lambda_\sigma \lesssim 55
\sqrt{H/T}$. Since warm inflation requires $T > H$ and in general we
typically work with values $T \gg H$, we find that weakly interacting
radiation fluids can easily have shear viscosities of sufficient
magnitude to counterbalance and suppress completely the growth of
fluctuations caused by the coupling of the inflaton's fluctuations with
those from the radiation.

\begin{figure}[t]
\vspace{0.5cm}
\includegraphics[width=0.55\textwidth,angle=0]{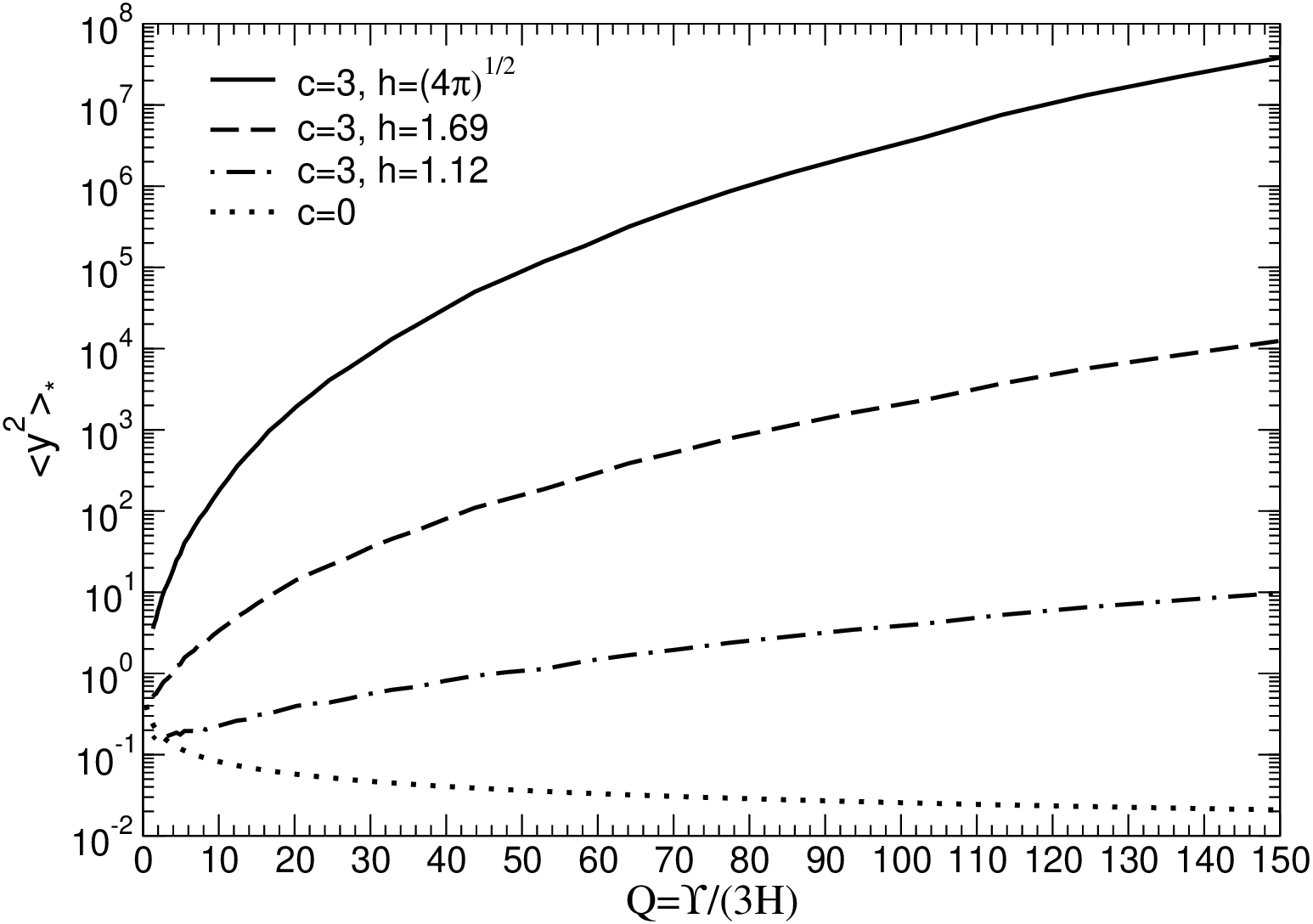}
\vspace{0.25cm}
\caption{\label{plot5} Field spectrum as a function of $Q_*$, for a
  quartic chaotic model, with a cubic dissipative coefficient
  (Eq. \ref{upslowT}) and $\zeta_s$ given by Eq. (\ref{zetaslf}). We
  have taken ${\cal N}_\sigma=10^3$ and $h=\sqrt{4\pi}$ (solid line),
  $h=1.67$ (dashed line) and $h=1.12$ (dot-dashed line). 
We include for comparison the result with $c=0$. 
} 
\end{figure}

In supersymmetric warm inflation models however the radiation bath
self coupling is the same as the coupling to the catalyst field, which
enters in the calculation of the dissipative coefficient. In general,
large multiplicities of the fields (catalyst $\chi$ and radiation $\sigma$)
are required in order to have enough dissipation. Considering
a model with ${\cal N}_\chi$, ${\cal N}_\sigma$ copies of the complex fields,
with common coupling $h$ among the $\chi$'s and $\sigma$'s, the
self-interaction potential is given by $h^2 \sum_i |\sigma|^2/4$. 
In the low-$T$ regime,  the dissipative and shear coefficients can be
written as \cite{BasteroGil:2010pb}:
\bea
\Upsilon & \simeq 0.1 h^4 {\cal N}_\chi{\cal N}_\sigma^2 \frac{T^3}{\phi^2}
\,, \label{upslowT}\\
\zeta_s &\simeq 127 N_\sigma (1 + 0.3 h)\frac{T^3}{h^4} \,, \label{zetaslf}
\end{align}
where we have included the next-to-leading order correction in the
shear viscosity \cite{Moore:2007ib}.  
The condition $\bar \zeta_s \geq 1$ then reads:
\be
h^4 \lesssim 128 \frac{{\cal N}_\sigma}{g_*} \frac{H}{T} \simeq 
69 \frac{H}{T} \,,
\ee
where $g_*\simeq 15 {\cal N}_\sigma/8$. Although 
in a bath of weakly interacting radiation particles, $h \ll 1$,   
the shear viscosity would be rather large, it may not give rise to  enough
dissipation to sustain a period of warm inflation, unless the weakness of
the coupling is compensated by having a large multiplicity for the fields.   
This therefore becomes a model dependent question depending on the
parameters ${\cal N}_\chi$, ${\cal N}_\sigma$ and the coupling
$h$. In Fig. \ref{plot5} we show an example for a quartic chaotic
model, where we compare the value of the field spectrum $\langle
y^2_k\rangle_*$ as a function
of $Q_*$ for different values of the Yukawa coupling $h$.  
We have kept the value ${\cal
  N}_\sigma=10^3$ fixed, and vary the value of ${\cal N}_\chi$ to get
different values of $Q_*$. From top to bottom the value of $h$
decreases, starting with the largest possible one $h=\sqrt{4\pi}$
(solid line) for which there is a negligible shear effect. In this case,  
a value $Q_*\simeq 10$ requires ${\cal N}_\chi=35 $, while for $Q_*\simeq
100$ we need ${\cal N}_\chi \simeq 110$. As the value of $h$
decreases, the multiplicity of the field to get the same value of
$Q_*$ increases by a factor $(4\pi/h^2)^2$.  For $h=1.12$ and $Q_*=100$,
we would need ${\cal N}_\chi \simeq 10^4$. 
As the value of $Q_*$ increases, so it does the
ratio $T/H$, and therefore the parameter $\bar \zeta_s$
decreases along the curves. Although  in this example 
viscous effects are not 
enough to completely avoid the growth of the perturbations, 
they bring  it down to $\langle y^2_k\rangle_* \propto Q^{2.5}$ for
$h=1,12$ and $\langle y^2_k\rangle_* \propto Q^{4.5}$ for $h=1.67$,
instead of $\langle y^2_k\rangle_* \propto Q^7$.      

\section{Conclusions}

Density perturbations in warm inflation are seeded by thermal
fluctuations of the inflaton. In warm inflation the inflaton decays into
radiation through a dissipation term in the inflaton's equation of
motion and that originates from the microphysical interactions of the
inflaton field with other degrees of freedom of the
microscopic Lagrangian describing the complete system. The origin of
the dissipation term and its quantum field theoretical treatment has
been extensively discussed in the literature (for a recent review, see
e.g. Ref. \cite{Berera:2008ar}).  However, as radiation is produced
during the inflaton's evolution, the full treatment of the spectrum of
perturbations no longer involves only that of the inflaton's
perturbations but must also account for radiation perturbations.  This
makes the treatment of density perturbations in warm inflation 
similar to
a multifluid system. Since the larger is the dissipation the larger is
expected to be the rate of radiation production, it becomes important
the study of how the produced radiation and its perturbations
backreacts on the inflaton's evolution and respective
perturbations. In Ref. \cite{mossgraham} it was shown that as
a consequence of this backreaction, the infaton's perturbations can
grow as the dissipation of the inflaton increases. This increasing of
the inflaton's perturbations with increasing dissipation can,
therefore, severely constrain the model parameters in warm inflation
so as to cope with the known measured results for the CMB
radiation. This backreaction of the produced radiation on the
inflaton's perturbations is larger the stronger is the coupling
between the radiation and the inflaton; in particular
the larger the power of $T$ in the dissipative coefficient,
the stronger the $T$ dependence on the perturbations.
This is exemplified by the results shown in
{}Fig. \ref{plot2}.
 
In this paper we have studied how this backreaction of the produced
radiation, that can lead to this growth mode in the inflaton's
perturbations, can be counterbalanced by the dissipative effects within
the radiation fluid. Dissipative effects in the radiation fluid
itself are described by viscosity terms.  This is expected when the
radiation fluid departs from equilibrium, which is the case in any
dissipative system, where the produced radiation from the system does
not immediately equilibrate in the radiation bath and its approach to
equilibrium is controlled by viscosity coefficients, like the shear
viscosity, the bulk viscosity and heat transport coefficients.  We
have focused on the dissipation effect as coming dominantly from a
shear viscosity term in the fluctuation equations. We have then shown
that the shear viscosity can effectively damp the radiation
fluctuations so as to avoid altogether the appearance of the growth mode
in the resulting perturbations. The results we have obtained are model
independent and we have shown that the overall effect of compensation
of the growth mode and its control is determined by the ratio $\zeta_s
H/(3 \rho_r)$, where $\zeta_s$ is the shear viscosity coefficient, $H$
is the Hubble parameter and $\rho_r$ the radiation energy density.
When the ratio is ${\cal O}(1)$ or larger, the growth mode disappears
completely.

In this work we have only considered the coupling between the
fluctuations in the inflaton field with those in 
the radiation through the
(temperature dependence on the) dissipation coefficient in the
inflaton's dynamics. But in a thermal bath, the parameters of the
inflaton's potential can also adquire temperature corrections. Even
though these thermal corrections can be kept under control and small
in Supersymmetry model building realizations for warm inflation
\cite{BR1,Berera:2008ar,BasteroGil:2009ec}, they can still be large enough
to provide extra sources of couplings between inflaton and radiation
fluctuations and it should be interesting to analyze their effects in a
future work. Likewise, there can be additional sources of dissipation
in the radiation fluid, for example as coming from bulk viscosity,
that can further help to damp any leftover growing modes as resulting from
these additional couplings. In this work we have neglected the effects
of the bulk viscosity on the grounds that it is in general much
smaller than the shear viscosity. {}For example, for the
self-interacting scalar field radiation discussed in section IV, the ratio of
the bulk viscosity, $\zeta_b$, with the shear viscosity for a high
temperature radiation fluid is $\zeta_b/\zeta_s\sim 10^{-9}
\lambda_\sigma^3$, thus, it is negligible for a weakly interacting
radiation bath. But there may be other interactions and energy
regimes for the radiation fluid in which the bulk can be sizable and lead
to effects in the density perturbation evolution (see
e.g. Ref. \cite{giovaninni}).

All these effects, starting with the shear viscosity, will also impact
the second order evolution of the perturbations and thus the
calculation of the non-gaussinity. Forthcoming cosmological data are
expected very soon to set the level of non-gaussinity of the
primordial spectrum, which clearly will help to discriminate among inflationary
models. Warm inflation, being a type of 
multi-fluid model, falls into the category of models with a
non-negligible value of the non-linearity parameter $f_{NL}$ for
non-gaussinity. This parameter has been computed for a $T$ independent
dissipative coefficient \cite{nongauss}, and recently the $T$
dependence of $\Upsilon$ has been included \cite{Moss:2011qc}, which
provides an extra 
non-linear source in the field second order equation. However, if the
coupling between field and radiation perturbations at first order is
suppressed by viscous effects, qualitatively we expect the same to
happen at second order. The question then is whether one simply
recover the prediction for a constant dissipative coefficient, or
non-linearities are further suppressed by viscous effects. 
These and other effects mentioned above will be studied
elsewhere.

\acknowledgements

A.B. acknowledges support from the STFC. R.O.R
is partially supported by Conselho Nacional de
Desenvolvimento Cient\'{\i}fico e Tecnol\'ogico (CNPq - Brasil).
M.B.G. would like to thank the hospitality of the School of 
Physics and Astronomy at the
University of Edinburgh which during her visit this work has started.
MBG is partially supported by MICINN (FIS2010-17395) and ``Junta de
Andaluc\'ia" (FQM101), and by SUPA during the realization of this work
in the UK.

\end{document}